\documentstyle[aps,epsfig]{revtex}
\begin{document}
\draft
\title{\hfill\begin{minipage}{0pt}\scriptsize \begin{tabbing}
\hspace*{\fill} Edinburgh-1999/10\\
\hspace*{\fill} SHEP-99/12\\
\hspace*{\fill} CERN/TH-99-265\\
\hspace*{\fill} CPT-99/P.3885\\
\hspace*{\fill} LAPTH-757/99\\
\hspace*{\fill} JLAB-THY-99-25\\
\end{tabbing}
\end{minipage}\\[8pt]
Improved $B\to\pi l\nu_l$ form factors from the lattice}
\author{UKQCD Collaboration}
\author{K.C.~Bowler$^{a)}$, L.~Del~Debbio$^{b)}$\footnote{Dipartimento di 
Fisica, Universit\`a di Pisa and INFN Sezione di Pisa, Italy.}
J.M.~Flynn$^{b)}$,
L.~Lellouch$^{c)}$\footnote{Present address, LAPTH
Chemin de Bellevue, B.P. 110, F-74941 Annecy-le-Vieux Cedex, France. 
On leave from CPT Luminy (UPR 7061),
Case 907, F-13288 Marseille, France}, V.~Lesk$^{b)}$,
C.M.~Maynard$^{a)}$, J.~Nieves$^{d)}$,
D.G.~Richards$^{a)}$\footnote{Jefferson
        Laboratory, MS 12H2, 12000 Jefferson Avenue, Newport News, VA 23606,
        USA}\\[5mm]}

\address{$^{a)}$Department of Physics \& Astronomy, University of Edinburgh,
        Edinburgh EH9 3JZ, Scotland, UK\\[0mm]
$^{b)}$Department of Physics \& Astronomy, University of Southampton,
        Southampton, SO17 1BJ, UK \\[0mm]
$^{c)}$Theory Division, CERN, CH 1211, Geneva 23, Switzerland\\[0mm]
$^{d)}$Departamento de F\'\i sica Moderna, Universidad de Granada, Granada 
18071,Spain}
\date{\today}
\maketitle
\begin{abstract}
We present the results of a lattice computation of the form factors
for $B^0\to\pi^- l^+\nu_l$ decays near zero-recoil.  These results
will allow a determination of the CKM matrix element $|V_{\rm ub}|$
when measurements of the differential decay rate become available.  We
also provide models for extrapolation of the form factors and rate to
the full recoil range. Our computation is performed in the quenched
approximation to QCD on a $24^3\times 48$ lattice at $\beta=6.2$,
using a non-perturbatively ${\mathcal O}(a)$-improved action. The
masses of all light valence quarks involved are extrapolated to their
physical values.
\end{abstract}


Semileptonic decays of heavy-light mesons, particularly those
containing a $b$ quark, have attracted considerable interest. They
play a crucial role in determining Cabibbo-Kobayashi-Maskawa (CKM)
matrix elements. With {\sc CLEO} and now {\sc BaBar} and Belle taking data, the
prospects for an accurate measurement of the differential rate for
$B\to\pi l\nu_l$ decays are excellent\footnote{CLEO already have a
measurement of the total rate~\protect\cite{CLEO_B2pi} and have been
able to obtain the differential decay rate for $B\to\rho l\nu_l$
decays in three recoil bins~\protect\cite{CLEObinnedrho}. They are
currently attempting a similar binning for $B\to\pi l\nu_l$
decays~\protect\cite{lawrence}.}. Combined with a first-principles
calculation of the relevant form factors, this measurement will allow
a model-independent extraction of $|V_{\rm ub}|$, currently one of
the least well-determined CKM parameters. 

The transition amplitude for exclusive semileptonic $b\to u$ decays
factorises into leptonic and hadronic parts. The hadronic matrix
elements contain the non-perturbative, strong-interaction effects and
are currently the largest source of theoretical uncertainty in the
determination of $|V_{\rm ub}|$ from these
decays~\cite{CLEO_B2pi,CLEObinnedrho}. We present here a lattice QCD
calculation in the quenched approximation of the matrix element
relevant for $B^0\to\pi^- l^+\nu_l$ decays. While this matrix element
has already been obtained from the lattice at restricted values of
$q^2$ \cite{Abada:1994dh} or for a range of $q^2$ with a
``pion'' composed of quarks with masses around that of the strange
quark \cite{ff_ukqcd_old}, we determine it for the
physical pion and a wider range of $q^2$ values~\footnote{Preliminary
  studies of these decays have already been presented by two
  groups~\protect\cite{chrism,chrism2,Ryan:1998tj,Ryan:1999er}.}. We perform a
fully ${\mathcal O}(a)$-improved calculation to reduce discretisation
errors and implement light quark mass extrapolations as suggested
in~\cite{lellouch}. The $q^2$-dependence of this matrix element has
also been calculated using light-cone sum rules \cite{lcsr} and a variety
of quark models \cite{qm}.

The matrix element can be parameterised in terms of two form factors:
\begin{eqnarray}
\label{eqn:form_factors}
  \langle \pi^-(\vec{k})| V^{\mu} | B^0(\vec{p}) \rangle
  &=& f_+(q^2)(p+k-q\Delta_{m^2})^{\mu} + f_0(q^2){}q^{\mu}\Delta_{m^2}
\end{eqnarray}
where $q=p-k$ and
  $\Delta_{m^2}=(M^2_B-m^2_{\pi})/q^2$.
The form factors $f_+$ and $f_0$ are both real, dimensionless functions
of the four-momentum transfer squared. In the limit of zero lepton mass the
differential decay rate is given by
\begin{equation}
\label{eqn:decay_rate}
  \frac{{\rm d}\Gamma}{{\rm d}q^2}
        =\frac{G_F^2\left |V_{\rm ub}\right|^2}{192\pi^3M_B^3}
    [\lambda(q^2)]^{3/2}\left |f_+(q^2)\right |^2,
\end{equation}
where $\lambda$ is a kinematic factor given by
\begin{equation}
  \lambda(q^2)=(M_B^2+m_{\pi}^2-q^2)^2-4M_B^2m_{\pi}^2
\end{equation}
and where $q^2$, the invariant mass-squared of the lepton pair, takes values
in the range from 0 to $(M_B-m_{\pi})^2$.

We work in the quenched approximation to QCD on a $24^3\times 48$
lattice at $\beta=6.2$, corresponding to an inverse lattice spacing of
$a^{-1}=2.54^{+5}_{-9}$ GeV as determined from the $\rho$ mass
\cite{ukqcd_prep}. We use 216 gauge configurations generated from a
combination of the over-relaxed \cite{creutz_or,brown_woch_or} and
Cabibbo-Marinari \cite{cabibbo_marinari} algorithms with periodic
boundary conditions.  The heavy-light mesons are composed of a
propagating heavy quark with mass around that of the charm quark to
keep discretisation errors in check, and a light antiquark with a mass
around that of the strange quark to keep finite volume errors and the
CPU time required to obtain the corresponding quark propagator under
control. We then obtain results for the physical $B^0\to\pi^- l^+\nu_l$
decay by extrapolating in heavy and light quark masses. We employ an
SW fermion action~\cite{sw_paper} that removes all ${\mathcal O}(a)$
discretisation errors from the hadron spectrum~\cite{alpha_np}. We can
also remove all ${\mathcal O}(a)$ discretisation errors from matrix
elements of local currents 
by an appropriate definition and renormalisation of these
currents~\cite{alpha_np}. In the case of the vector current, we have
\begin{equation}
  V^I_{\mu} = V_{\mu}^L+c_Va
        \frac{1}{2}({\partial}_{\nu}+{\partial}_{\nu}^{\star})T_{\mu\nu}^L
\end{equation}
where $V_{\mu}^L=\bar Q\gamma_\mu q$ and $T_{\mu\nu}^L=\bar
Q\sigma_{\mu\nu}q$ are the local lattice vector and tensor currents
respectively and where ${\partial}_{\nu}$
(${\partial}_{\nu}^{\star}$) is the forward (backward) lattice derivative. 
The renormalised current is
\begin{equation}
  V_{\mu} = Z_V\left(1 + \frac{b_V}{2} (am_Q+am_q)\right)V^I_{\mu}\ 
	+ {\mathcal O}(a^2)\ ,
\end{equation}
where $am_{Q,q}=1/2\kappa_{Q,q} - 1/2\kappa_{\rm crit}$ and both
$b_V$ and the matching coefficient $Z_V$ have been determined
non-perturbatively~\cite{alpha_np3_Z}. The mixing coefficient $c_V$ is
only known to one loop in perturbation theory~\cite{alpha_np4_bA}, but
is small.

To quantify residual discretisation errors, we consider the effective
matching coefficient, $Z_V^{\rm eff}$, obtained from the forward
matrix element of the current, $V_{\mu}=\bar Q\gamma_\mu Q$, between
degenerate heavy-light mesons at rest. Since, in the continuum, this
current is conserved we have
\begin{equation}
\label{eqn:ZV_eff}
  \langle P(\vec{0})\; | V_0 | P(\vec{0})\; \rangle
	= 2 M_P  = Z_V^{\rm eff} 
     \langle P(\vec{0}) \;| V^I_0 | P(\vec{0}) \;\rangle 
\end{equation}
where $P$ is a heavy-light pseudoscalar meson. 
We calculate $Z_V^{\rm eff}$ for two
values of the heavy quark mass. We compare these results with the
matching coefficient, $Z_V(1+b_Vam_Q)$, evaluated for these quark
masses using the non-perturbative values of the coefficients $Z_V$ and
$b_V$. These two measures will differ by terms of
${\mathcal O}(a^2)$. The comparison is shown in Table
\ref{tab:zv_eff}. The agreement between the two procedures is
excellent, suggesting that, at least for short-distance quantities,
higher-order discretisation errors are small.  However, we cannot
exclude the possibility that discretisation errors for non-degenerate,
non-forward matrix elements are larger than the above comparison
suggests.  Although the present calculation has been performed at one
value of the lattice spacing, recent work~\cite{garden_strange} using
the same improved action, suggests that for long-distance quantities
such as masses the error in extrapolating to the continuum limit
from $\beta=6.2$ is around $6\%$.

Four values of the hopping parameter, $\kappa$, were chosen for the
heavy quark ($\kappa_Q=0.1200$, $0.1233$, $0.1266$, $0.1299$), three values
for the active light quark (A) ($\kappa_A=0.1346$, $0.1351$, $0.1353$),
and two for the spectator light quark (S) ($\kappa_S=0.1346$, $0.1351$).
We obtain the relevant matrix elements from heavy-to-light
three-point correlation functions, divided by the appropriate factors
extracted from two-point functions as in
Refs.~\cite{bowler_d2k,bowler_b2kstar_gamma}. We place the operator
for the heavy-light pseudoscalar at $t=20$, and the operator for the
light-light pseudoscalar at $t = 0$. We find that
contamination of the signal from other time orderings as well as from
excited states is negligible.

We use eight different combinations of $\vec{p}$ and $\vec{k}$ to
determine the momentum dependence of the corresponding form factors:
$0\to0$, $0\to1$, $0\to\sqrt 2$, $1\to0$, $1\to1$, $1\to1_{\bot}$,
$1\to1_{\gets}$ and $1\to\sqrt 2_{\bot}$ in lattice units. There is no
$0\to0$ channel for $f_+$.  For each momentum channel, the form
factors are obtained from a simultaneous fit to the matrix elements of
the temporal and spatial components of the vector current.

The chiral limit is reached at the value of the hopping parameter for
which the pion mass vanishes. For the present simulation this value is $\kappa_{\rm
crit}=0.13582(1)$. The physical value
of $m_{\pi}/m_{\rho}$ is reached for $\kappa_n=0.13578(1)$. In order to 
evaluate the form factor, $f_i$, $i=+,\ 0$, at physical light quark masses, we
must consider both the explicit mass dependence of $f_i$ and
the indirect dependence arising from the change in $q^2$:
\begin{equation}
f_i=f_i(q^2_{(\kappa_A,\kappa_S)},\kappa_A,\kappa_S).
\end{equation}

\begin{figure}
\begin{center}
\caption{Interpolations of $f_+(q^2)$ and $f_0(q^2)$ for the heaviest of the 
        heavy quarks and the lightest of the light quarks. The 
        support of the curves shows the range of $q^2$ which results in 
        interpolation for all light quark mass combinations. The fits shown
        enforce the kinematic constraint $f_+(0)=f_0(0)$.}
\vspace{1em}
\epsfig{file=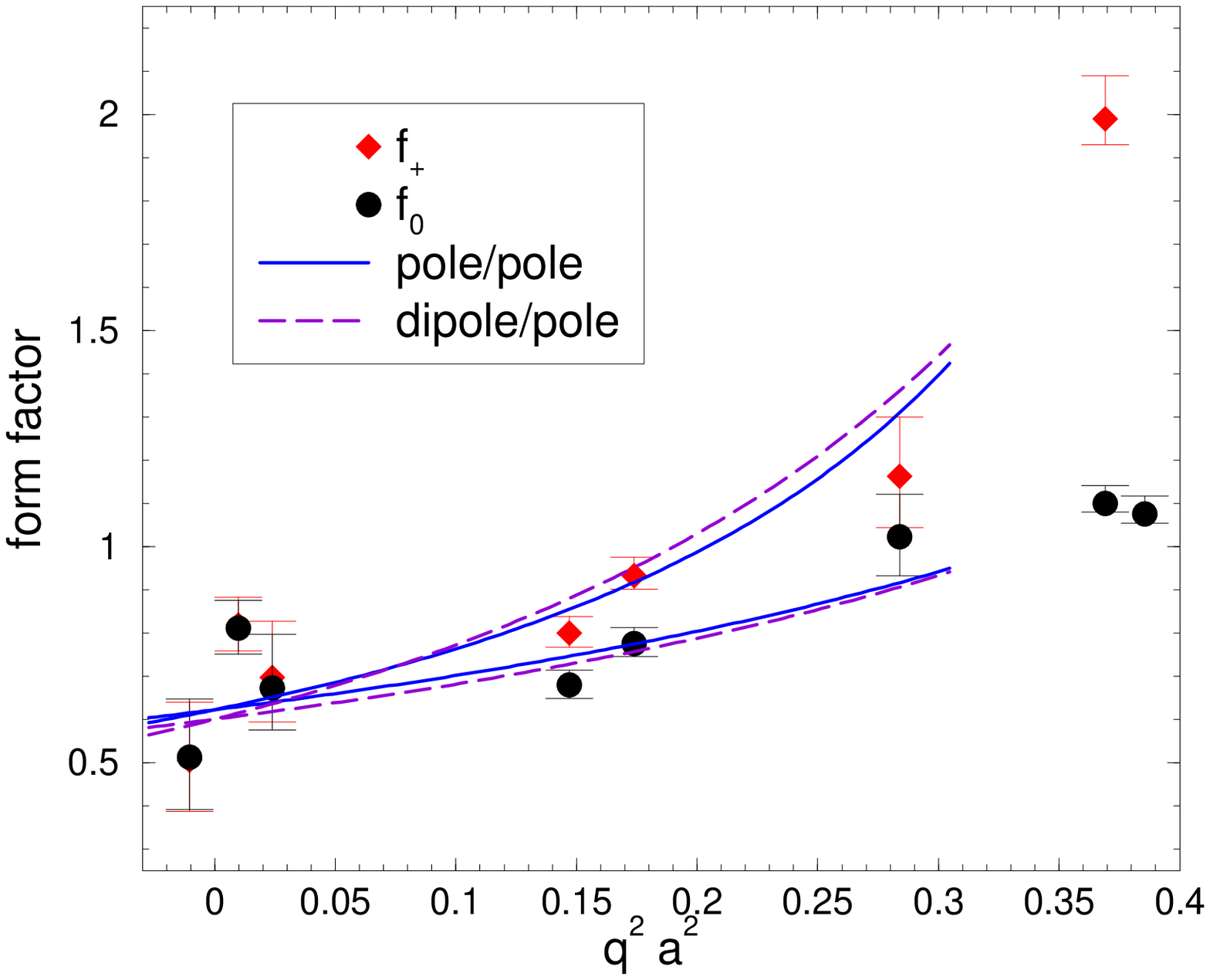,height=3.375in,width=3.375in}
\label{fig:fixqsqr}
\end{center}
\end{figure}

We first interpolate, to a common set of $q^2$ values, the form
factors corresponding to a given heavy quark and different
light quark mass combinations. For the different heavy quarks, these
sets are chosen such that the corresponding sets of heavy quark
velocities are the same for all heavy quarks, for
reasons that will become clear below. The interpolation is performed
at fixed light quark mass using a form motivated by pole dominance
models:
\begin{equation}
  f_i(q^2)=\frac{f_i(0)}{1-q^2/M_i^2}\ ,
\label{eq:pd}
\end{equation}
with the kinematical constraint, $f_+(0)=f_0(0)$, imposed.  Any
consequent model dependence is mild as we use this ansatz only to
interpolate the form factors in the range of $q^2$ for which we have
data. We have also tried other ans\"{a}tze including dipole/pole 
for $f_+(q^2)$ and $f_0(q^2)$ respectively, with and/or
without the kinematical constraint. This is included in our estimate
of systematic errors and is shown in Figure \ref{fig:fixqsqr}.

These interpolated data points are then extrapolated in $\kappa_A$
and $\kappa_S$ to $\kappa_n$ at {\itshape fixed $q^2$}, according to
\begin{equation}
\label{eqn:chiral_FF_k}
  f(\kappa_{S}, \kappa_{A})=\alpha + \beta \left(
  \frac{1}{\kappa_S}-\frac{1}{\kappa_{\rm crit}} \right )
  +\gamma\left(\frac{1}{\kappa_S}+\frac{1}{\kappa_A}-\frac{2}{\kappa_{\rm crit}}\right
  ).
\end{equation}
The extrapolation is
two-dimensional in the active and spectator light quark masses.

In previous UKQCD collaboration analyses~\cite{chrism,bowler_d2k} a
term linear in the pion mass was added to account for the indirect
$q^2$ contribution. Here
we hold $q^2$ fixed as the light quark masses are changed~\cite{lellouch}. 
This yields a more reliable extrapolation by separating the $q^2$ and the
explicit light quark mass dependences.

Heavy quark symmetry (HQS) is used to model the form factors'
dependence on heavy meson mass at fixed four-velocity, $v$. We thus
work with the recoil variable
\begin{equation}
  v\cdot k= \frac{M_P^2 + m_{\pi}^2 - q^2}{2 M_P}.
\end{equation}
The scaling relations for $f_+$ and $f_0$
at fixed $v\cdot k$, as given by heavy quark effective theory 
are~\cite{H_scaling},
\begin{equation}
\label{eqn:heavy_extrap}
  C(M_P,M_B)f_i(v\cdot k)M_P^{s_i/2}=
        \gamma_i\left (1 +\frac{\delta_i}{M_P}
         + \frac{\epsilon_i}{M^2_P} + \cdots \right)\
\end{equation}
where the ellipsis denotes higher order terms in the heavy quark expansion
and $s_i=\{-1,1\}$ when $i=\{+,0\}$. The coefficient $C$ is
the logarithmic matching factor~\cite{neubert_coeff},
\begin{equation}
\label{eqn:HQET_match}
  C(M_P,M_B)=\left (\frac{\alpha_s(M_B)}{\alpha_s(M_P)}\right)
        ^{2/\beta_0}
\end{equation}
where $\beta_0=11$ in quenched QCD and $\alpha_s$ is the one
loop running coupling with $\Lambda^{(4)}_{\overline{\rm MS}}=295$ MeV.

\begin{figure}
\begin{center}
\caption{Heavy-quark-mass extrapolation of $f_+(q^2)$ at fixed $v\cdot k$ 
	corresponding to $q^2=22.3$ GeV$^2$ at the B meson scale.}
\vspace{1em}
\epsfig{file=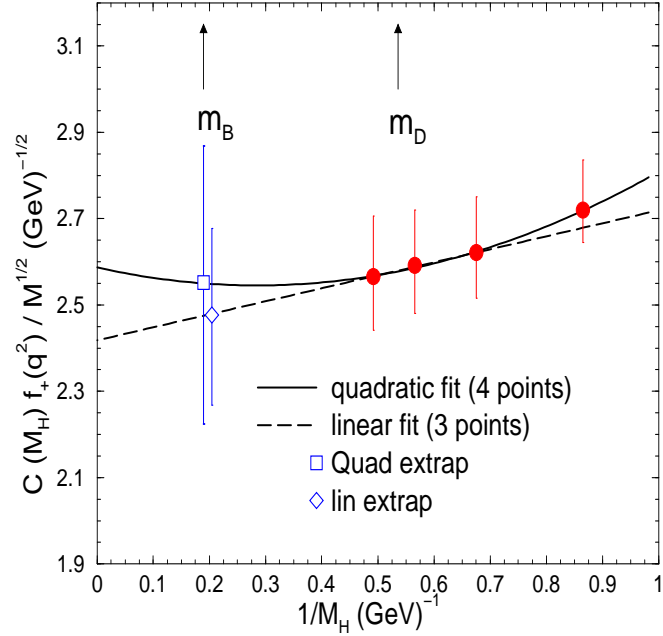,height=3.375in,width=3.375in}
\label{fig:HQS}
\end{center}
\end{figure}

We choose a set of five values of $v\cdot k$ in a region accessible to
our calculation for all of our values of heavy and light quark
mass. For each initial heavy quark this yields a distinct set of $q^2$
values. It is these sets of $q^2$ values which we use in the light
quark mass extrapolation described above.  From that extrapolation, we
obtain the form factors for all $M_P$ at the chosen values of $v\cdot
k$.  We then extrapolate $f_+$ and $f_0$ at fixed $v\cdot k$ to obtain
the form-factors at the $B$-meson mass.  Figure \ref{fig:HQS} shows
both a quadratic fit to all four data points and a linear fit to the
three heaviest. The difference between these two procedures is used to
estimate the systematic uncertainty in this extrapolation. 
Table~\ref{tab:quark_mass} shows the heavy quark and meson masses used
in this work.

\begin{figure}
\begin{center}
\caption{Momentum dependence of the form factors. The data shows statistical
    errors only.}
\vspace{1em}
\epsfig{file=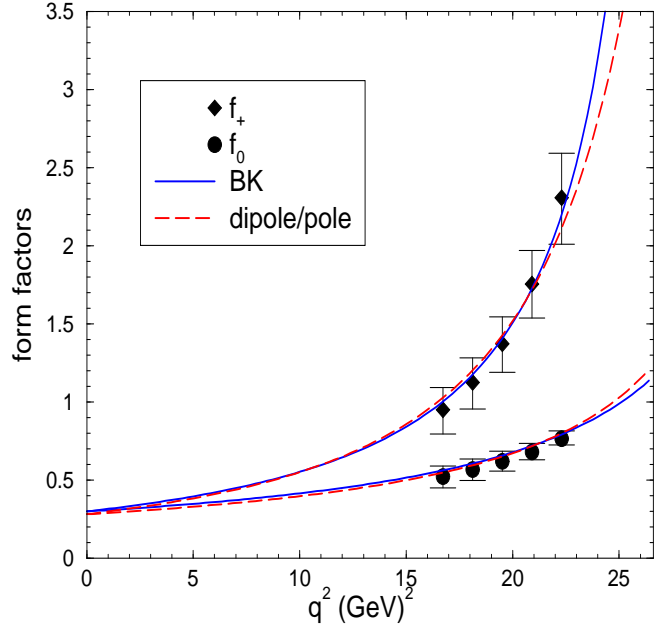,height=3.375in,width=3.375in}
\label{fig:B2PI_con}
\end{center}
\end{figure}

\begin{figure}
\begin{center}
\caption{The differential decay rate as a function of $q^2$. The outer error
        bars show the systematic and statistical errors added in quadrature.
        The curves are model fits to both $f_+$ and $f_0$.}
\vspace{1em}
\epsfig{file=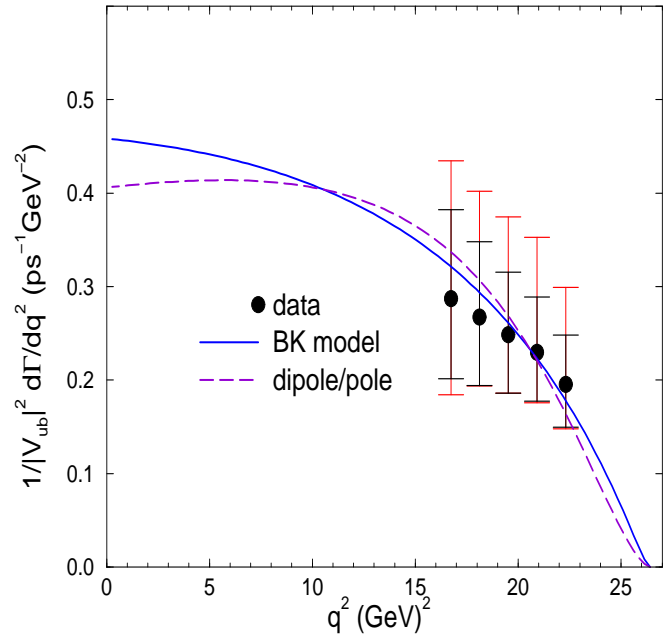,height=3.375in,width=3.375in}
\label{fig:diff_rate}
\end{center}
\end{figure}

The resulting $B^0\to\pi^- l^+\nu_l$ form factors and differential
decay rate are shown as functions of $q^2$ in Figures
\ref{fig:B2PI_con} and \ref{fig:diff_rate} and their values are
summarised in Table \ref{tab:diff_rate}. It is important to note that
these results can be used for a determination of $|V_{\rm ub}|$,
without model-dependent assumptions about the $q^2$ dependence of form
factors, once experiments measure the differential or partially
integrated rate in the range of $q^2$ values reached by our
calculation. Future lattice calculations in full, unquenched QCD will
permit a completely model-independent determination of $|V_{\rm ub}|$.

The central values and statistical, bootstrap errors in the figures
and table are obtained with the kinematically constrained pole/pole
$q^2$-interpolation function of Eq.~(\ref{eq:pd}), quadratic
heavy-quark-mass extrapolations and with the lattice spacing set by
$m_{\rho}$.  Systematic errors are determined by considering the
change in our results due to: different choices of $q^2$-interpolation
function, as discussed after Eq.~(\ref{eq:pd}); using
$r_0$~\cite{wittig_r0} to set the lattice spacing; and using a linear,
instead of quadratic, heavy extrapolation for the heaviest three
quarks.  Using the KLM quark field normalisation~\cite{KLM} produced
negligible change in these results, suggesting that higher order
mass-dependent discretisation effects are small. We also assign a
$6\%$ error representing the possible effect of a continuum
extrapolation as mentioned above. The quoted error is the quadratic
sum of the maximum variation from all sources of systematics.

To extrapolate these results to a larger range of $q^2$, we 
make model assumptions~\footnote{In \protect\cite{lellouch}, dispersive bound
techniques were used to extrapolate lattice results to the full kinematical
range in a model-independent way. Such an extrapolation, however, 
is beyond the scope of the present
letter.}. Pole dominance models suggest the following
momentum dependence for the form factors,
\begin{equation}
  f_i(q^2)=\frac{f(0)}{(1-q^2/M_i^2)^{n_i}},
\end{equation}
where $i=+,\ 0$, $n_i$ is an integer exponent and the kinematical
constraint $f_+(0)=f_0(0)$ has already been imposed.  Combining this
with the HQS scaling relations of Eq.~(\ref{eqn:heavy_extrap})
implies $n_+=n_0+1$. Light-cone sum-rules scaling further suggests
$n_0=1$~\cite{DelDebbio:1997kr}~\footnote{Pole/dipole behaviour for $f_0$ and 
$f_+$ was also found in \cite{Charles:1998dr}, where the scaling of 
heavy-to-light form factors with initial heavy meson mass and final light meson
energy (in the heavy meson rest frame) was investigated systematically.}. 
Another pole/dipole parameterisation for $f_0$ and $f_+$, which
accounts for the $B^*$ pole in $f_+$ correctly, has been suggested by
Becirevic and Kaidalov
(BK)~\cite{BK_param}:
\begin{eqnarray}
  f_+(q^2)&=&\frac{c_B(1-\alpha)}
  {(1-q^2/M^2_{B^{\star}})(1-\alpha q^2/M^2_{B^{\star}})} \nonumber \\
  f_0(q^2)&=&\frac{c_B(1-\alpha)}{(1- q^2/\beta M^2_{B^{\star}})}.      
\end{eqnarray}

We fit both models to our form factors extrapolated to the $B$-meson
mass. The results are shown in Figure \ref{fig:B2PI_con} for our
central value procedure. Though both parameterisations fit the data
equally well, we favour the more physical BK description.

These models can then be used to compute the total decay rate by
integrating Eq.~(\ref{eqn:decay_rate}) with respect to $q^2$. This
rate of course depends on the choice of model and is sensitive to
uncertainties in the lattice results, since it is dominated by the
low $q^2$ region, only reached by extrapolation. We find
\begin{equation}
\label{eqn:my_decay_rate}
  \Gamma/|V_{\rm ub}|^2=\left (9^{+3}_{-2}{}^{+2}_{-2}\right )\ {\rm ps}^{-1}.
\end{equation}
The first error is statistical and the second is the systematic error
estimated, as for the form factors, from the effects decribed above as
well as the difference between the two models. Combining this result
with CLEO's exclusive $|V_{\rm ub}|=(3.25\pm0.14^{+0.21}_{-0.29}\pm
0.55)\times 10^{-3}$~\cite{CLEObinnedrho} and the $B^0$ lifetime,
$\tau_{B^0}=1.54\pm 0.03\ {\rm ps}$~\cite{pdg98}, we find a branching ratio
$\Gamma/\Gamma_{\rm total}=\left (1.5^{+0.5}_{-0.3}{}^{+0.3}_{-0.3}\pm0.6\right
) \times10^{-4}$, where the first error is the lattice statistical
error, the second systematic and the third, the experimental errors on
the mean combined in quadrature. This result is consistent with the
measurement by the CLEO collaboration \cite{CLEO_B2pi},
$\Gamma/\Gamma_{\rm total}=(1.8\pm0.4\pm0.3\pm0.2)\times10^{-4}$,
where the errors are statistical, systematic and from model
dependence.

In this letter we have reported a lattice computation of the form
factors and differential decay rate for the decay $B^0\to\pi^- l^+\nu_l$.  We have used a
fully ${\mathcal  O}(a)$-improved action to minimise discretisation errors.
We are repeating the calculation at a different lattice spacing to probe
the effect of improvement. It should also be remembered that the
computation was performed in the quenched approximation; we are
currently generating configurations with two flavours of light
dynamical quarks and aim to quantify the quenching effect.

We acknowledge EPSRC grant GR/K41663, and PPARC grants
GR/L29927 and GR/L56336.  DGR acknowledges PPARC, and the DOE under
contract DE-AC05-84ER40150, and thanks FNAL for their hospitality
during part of this work. We thank the British Council and Spanish DGES 
for travel support under the Acciones Integradas scheme 1998/99, grants 1786 
and HB1997-0122. JN acknowledges the Spanish DGES for support under contract 
PB95-1204. We are grateful for helpful correspondence and discussions with 
Lawrence Gibbons and Steve Playfer.~\cite{alpha_np4_bA,alpha_ZM}

\pagebreak
\pagebreak
\pagebreak
\begin{table}
\caption{The effective matching coefficient for different values of the heavy 
        quark mass and spectator hopping parameter. The current does not depend
        on the spectator quark.}
\label{tab:zv_eff}
\begin{tabular}{cccc}
$am_Q$& $Z_V^{\rm eff},\kappa_S=0.1346$ & $Z_V^{\rm eff},\kappa_S=0.1351$ &
        $Z_V(1+b_Vam_Q)$ \\
$0.485$& $1.315^{+3}_{-3}$& $1.316^{+4}_{-5}$ &$1.332^{+4}_{-4}$\\
$0.268$& $1.087^{+2}_{-2}$&$1.085^{+3}_{-2}$&$1.090^{+3}_{-3}$
\end{tabular}
\end{table}

\begin{table}
\begin{center}
\caption{Heavy quark and meson masses used in this work. The table shows
	the {\em bare} quark mass in lattice units, 
	the {\em renormalisation group invariant} quark mass defined by 
	$m^{\rm RI}_Q=Z_M(1+b_Mam_Q)m_Q$\protect\cite{alpha_ZM}
	and the heavy-light($\kappa_n$) pseudoscalar meson mass. The scale is
	set by $m_\rho$ and $b_m$ is obtained using the one-loop result
	of~\protect\cite{alpha_np4_bA}}
\label{tab:quark_mass}
\begin{tabular}{cccc}
$\kappa$ & $am_Q$ & ${m}^{\rm RI}_Q$ (GeV) & $M_P$ (GeV)\\
\hline
0.1200  & 0.485    & 1.52 & $2.035\pm5$\\
0.1233  & 0.374   & 1.23 & $1.771\pm5$ \\
0.1266  & 0.268   & 1.02 & $1.483\pm5$\\
0.1299  & 0.168   & 0.69 & $1.157\pm5$\\
\end{tabular}
\end{center}
\end{table}

\begin{table}
\begin{center}
\caption{Form factors and differential decay rate as functions of $q^2$.
        The central value comes from the constrained pole/pole interpolation 
	procedure, the first error is statistical, the second is systematic as
	described in the text.}
\label{tab:diff_rate}
\begin{tabular}{cccccc}{}
$q^2$ $($GeV$)^2$ & $16.7$ & $18.1$ & $19.5$ & $20.9$ & $22.3$ \\
$f_+(q^2)$ & $0.9^{+1}_{-2}\ {}^{+2}_{-1}$ 
        & $1.1^{+2}_{-2}\ {}^{+2}_{-1}$ 
        & $1.4^{+2}_{-2}\ {}^{+3}_{-1}$ 
        &$1.8^{+2}_{-2}\ {}^{+4}_{-1}$
        & $2.3^{+3}_{-3}\ {}^{+6}_{-2}$ \\     
$f_0(q^2)$ & $0.57^{+6}_{-6}\ {}^{+\ 6}_{-20}$
        & $0.61^{+6}_{-6}\ {}^{+\ 7}_{-19}$ 
        & $0.66^{+5}_{-5}\ {}^{+\ 7}_{-17}$ 
        & $0.72^{+5}_{-4}\ {}^{+\ 7}_{-15}$ 
        & $0.79^{+5}_{-4}\ {}^{+\ 8}_{-12}$ \\
$1/|V_{\rm ub}|^2{\rm d}\Gamma / {\rm d}q^2$ $({\rm ps}^{-1}{\rm GeV}^{-2})$
        & $0.29^{+10}_{-\ 9}\ {}^{+11}_{-\ 6}$
        & $0.27^{+8}_{-7}\ {}^{+11}_{-\ 2}$ 
        & $0.25^{+7}_{-6}\ {}^{+11}_{-\ 2}$
        & $0.23^{+6}_{-5}\ {}^{+11}_{-\ 2}$ 
        & $0.20^{+5}_{-5}\ {}^{+\ 9}_{-\ 2}$
\end{tabular}
\end{center}
\end{table}
\end{document}